\documentstyle[twoside,amssymb,12pt]{article}

\newcommand{\nc}{\newcommand}
\nc{\la}{\lambda} \nc{\alf}{\alpha}  \nc{\T}{\Theta}
\nc{\tht}{\theta}  \nc{\be}{\beta}  \nc{\eps}{\epsilon} \nc{\ga}{\gamma}  
\nc{\De}{\Delta}  \nc{\Ga}{\Gamma}  \nc{\vphi}{\varphi}  \nc{\z}{\zeta}
\nc{\de}{\delta} \nc{\si}{\sigma}  \nc{\ka}{\kappa}   \nc{\Si}{\Sigma}
\nc{\om}{\omega}  \nc{\qq}{\quad\quad}                \nc{\Om}{\Omega}
\nc{\nf}{\infty}   \nc{\dl}{\mathop{\smash{\cal L}}}  \nc{\black}{\rule{3mm}{3mm}}
\nc{\ra}{\rightarrow}  \nc{\ol}{\overline}  \nc{\und}{\underline}
\nc{\beq}{\begin{equation}}  \nc{\pt}{\partial}
\nc{\eeq}{\end{equation}}
\nc{\beqa}{\begin{eqnarray}}  \nc{\dst}{\displaystyle}
\nc{\eeqa}{\end{eqnarray}} \nc{\nnb}{\nonumber}  \nc{\ti}{\tilde}
\nc{\bs}{\backslash}        \nc{\mb}{\mathbb}  \nc{\wti}{\widetilde} \nc{\wh}{\widehat}

\newcounter{muni}
\newenvironment{remunerate}{\begin{list}{{\rm \arabic{muni}.}}
{\usecounter{muni}
\setlength{\leftmargin}{0pt}\setlength{\itemindent}{38pt}}}{\end{list}}

\nc{\brm}{\begin{remunerate}}   \nc{\erm}{\end{remunerate}}

\newtheorem{nth}{Proposition}   \newtheorem{nlem}{Lemma}  
\nc{\stg}{\mathop{\smash{*}}}     \nc{\st}{\mathop{\smash{\delta}}}
\nc{\stp}{\mathop{\smash{\otimes}}}
\nc{\barr}{\begin{array}}   \nc{\earr}{\end{array}}   \nc{\dg}{\dagger}
\nc{\mtvb}{\mathversion{bold}}   \nc{\mtvn}{\mathversion{normal}}

\date{}
\begin{document}\begin{titlepage}
\begin{flushright}
hep-th/0309207 \\
September 2003
\end{flushright} 
\vskip 2.0truecm
\centerline{\large \bf INTEGRABILITY VERSUS SEPARABILITY}
\centerline{\large \bf FOR THE MULTI-CENTRE METRICS}
\vskip 1.0truecm 
\centerline{\bf Galliano Valent${}^{*\dagger}$}
\vskip 1.0truecm 

\centerline{${}^{*}$\it Laboratoire de Physique Th\'eorique et des
Hautes Energies}
\centerline{\it Unit\'e associ\'ee au CNRS UMR 7589}
\centerline{\it 2 Place Jussieu, 75251 Paris Cedex 05, France} 
\vskip 0.2truecm

\vspace{5mm}\centerline{${}^{\dagger}$\it CNRS Luminy Case 907 }
\centerline{\it Centre de Physique Th\'eorique}
\centerline{\it F-13288 Marseille Cedex 9}
\nopagebreak
\vskip 0.5truecm

\begin{abstract}
The multi-centre metrics are a family of euclidean solutions of the empty space 
Einstein equations with self-dual curvature. For this full class, we determine 
which metrics do exhibit an extra conserved quantity quadratic in the momenta, 
induced by a Killing-St\" ackel tensor. Our systematic approach brings to light 
a subclass of metrics which correspond to new classically integrable dynamical 
systems. Within this subclass we analyze on the one hand the separation 
of coordinates in the Hamilton-Jacobi equation and on the other hand the 
construction of some new Killing-Yano tensors.
\end{abstract}
\end{titlepage}

\section{Introduction}
The discovery of the generalized Runge-Lenz vector for the Taub-NUT  
metric \cite{gm} has been playing an essential role in the analysis of its 
classical and quantum dynamics. As shown in \cite{fh} this triplet 
of conserved quantities gives quite elegantly the quantum bound states 
as well as the scattering states.  

The Killing-St\" ackel tensors, which are the roots of the 
generalized Runge-Lenz vector of Taub-NUT, have been derived in \cite{gr} using 
purely geometric tools. As a result the classical integrability 
of the Taub-NUT metric was established. The classical integrability of the 
Eguchi-Hanson metric was obtained in \cite{Mi} where the Hamilton-Jacobi 
equation was separated. This result was further generalized in \cite{gr} to cover the 
two-centre metric. Despite these successes, a systematic analysis of the full 
family of the multi-centre metrics was still lacking. It is the aim of this 
article to fill this gap.

In section 2 we have gathered a summary of known properties of the 
multi-centre metrics, their geodesic flow and some basic concepts about 
Killing-St\" ackel tensors.

In section 3 we obtain the most general structure of the conserved quantity 
associated to a Killing-St\" ackel tensor: it is a bilinear form in the momenta. 
Taking this quadratic structure as a starting point, we obtain the 
system of equations which ensure 
that such kind of a quantity is preserved by the geodesic flow. This system is 
analyzed and simplified. Its most important consequence is that the 
existence of an extra conserved quantity is related to the existence of an 
extra spatial Killing (besides the tri-holomorphic one), which 
may be either holomorphic or tri-holomorphic.

In section 3 we first consider the case of an extra spatial Killing which 
is holomorphic. We find that the extra conserved quantity does exist for the 
following families, with (minimal) isometry $U(1)\times U(1)$:
\brm
\item The most general two-centre metric, with the potential
\[V=v_0+\frac{m_1}{|\vec{r}+\vec{c}|}+\frac{m_2}{|\vec{r}-\vec{c}|}.\]
Our approach explains quite simply why there are three extra conserved 
quantities for Taub-NUT and only one for Eguchi-Hanson, and their 
very different nature.
\item A first dipolar breaking of Taub-NUT, with potential
\[V=v_0+\frac mr+\frac{\vec{\cal F}\cdot\vec{r}}{r^3}.\]
\item A second dipolar breaking of Taub-NUT with potential
\[V=v_0+\frac mr+\vec{\cal E}\cdot \vec{r}.\]
In the Taub-NUT limit ${\cal E}\to 0$ there appears a triplet of extra 
conserved quantities: the generalized Runge-Lenz vector of \cite{gm}.

The classical integrability of these three dynamical systems follows from 
our analysis.
\erm

In section 4 we consider the case of an extra spatial Killing which is 
tri-holomorphic, with (minimal) isometry group still $\,U(1)\times U(1).$ 
We find four different families of metrics, which share with the previous 
ones their classical integrability and, using appropriate coordinates, 
with potentials:
\brm
\item  In the first case
\[V=v_0+\frac{a\xi\sqrt{\xi^2-c^2}+b\eta\sqrt{c^2-\eta^2}}{\xi^2-\eta^2}.\]
\item In the second case
\[V=v_0+m\,\frac{\cos(2\phi)}{r^2}.\]
\item In the third case
\[V=\frac{a\xi+b\eta}{\xi^2+\eta^2}.\]
\item And in the fourth case
\[V=v_0+mx.\]
\erm
 
As an application we work out in sections 5 and 6 the separation of variables 
for the Hamilton-Jacobi equation which gives also a check of the results 
obtained in the former sections.

Eventually we present in section 7 some new  Killing-Yano tensors, and some 
conclusions in section 8.

\section{The Multi-Centre metrics}
\subsection{Background material}
These euclidean metrics on $M_4$ have at least one 
Killing vector $\,\wti{\cal K}=\pt_t\,$ and have the local form
\beq\label{mc1}
g=\frac 1V\,(dt+\T)^2+V\,\ga,\qq V=V(x),\qq\T=\T_i(x)\,dx^i,
\eeq
where the $x^i$ are the coordinates on $\ga.$ They are solutions of the empty 
space Einstein equations provided that :
\brm
\item The three dimensional metric $\ga$ is flat. Using cartesian coordinates $x^i$ 
we can write
\beq\label{mc2}
\ga=d\vec{x}\cdot d\vec{x}.\eeq
\item Some monopole equation holds
\beq\label{mc3}
dV=\,\stg_{\ga}\,d\T.\eeq
\erm
Notice that the integrability condition for the monopole equation is $\,\De V=0,$ 
hence these metrics display an exact linearization of the empty space Einstein 
equations. 
They have been derived in many ways \cite{ksd},\cite{gh},\cite{Hi1},\cite{Hi2}. 
In this last reference the geometric meaning of the cartesian coordinates $x_i$ was 
obtained: they are nothing but the momentum maps of the complex 
structures under the circle action of $\pt_t.$

Let us summarize some background knowledge on the multi-centre metrics for further 
use. Taking for canonical vierbein
\beq\label{fv}
E_a\ :\qq\quad E_0=\frac 1{\sqrt{V}}(dt+\T),\qq\qq E_i=\sqrt{V}\,dx_i\eeq
and defining as usual the spin connection $\,\om_{ab}\,$ and the 
curvature $\,R_{ab}\,$ by
\[d E_a+\om_{ab}\wedge E_b=0,\qq\quad R_{ab}=d\om_{ab}+\om_{as}\wedge \om_{sb},\]
one can check that these metrics have a self-dual spin connection: 
\[\om^{(-)}_i\equiv \om_{0i}-\frac 1{2}\,\eps_{ijk}\,\om_{jk}=0,\qq
\Longrightarrow\qq R^{(-)}_i=0,\] 
which implies the self-duality of their curvature. It follows that they 
are hyperk\" ahler and hence Ricci-flat.

The complex structures are given by the triplet of 2-forms
\beq\label{f7}
\Om^{(-)}_i=E_0\wedge E_i-\frac 1{2}\,\eps_{ijk}\,E_j\wedge E_k=(dt+\T)\wedge dx_i-
\frac 1{2}\, V\,\eps_{ijk}\,dx_j\wedge dx_k, \eeq
which are closed, in view of the hyperk\" ahler property of these metrics. 

Let us note that the self-duality of the complex structures and of the spin 
connection are opposite and that the Killing vector $\,\pt_t\,$ is tri-holomorphic. 

It is useful to define the Killing 1-form, dual of the vector $\wti{\cal K}=\pt_t,$ 
which reads 
\beq\label{f8}
{\cal K}=\dst\frac{dt+\T}{V},\eeq
and plays some role in characterizing the multi-centre metrics. 

\noindent Among these characterizations let us mention:
\brm
\item For the multi-centre metrics the differential $d{\cal K}$ has a self-duality 
opposite to that of the connection. A proof using spinors may be found in \cite{tw} 
and without spinors in \cite{gd}.
\item The multi-centre metrics possess at least one tri-holomorphic Killing. For 
a proof see \cite{gr}.
\erm

\nc{\dt}{\dot{t}}   \nc{\dx}{\dot{x}}    \nc{\dy}{\dot{y}}    \nc{\dep}{\dot{p}}

\subsection{Geodesic flow}
The geodesic flow is the Hamiltonian flow of the metric considered as a function on 
the cotangent bundle of $M_4.$
Using the coordinates $(t,x_i)$ we will write a cotangent vector as
\[\Pi_i\,dx_i+\Pi_0\,dt.\]
The symplectic form is then
\beq\label{gf1}
\om=dx_i\wedge d\Pi_i+dt\wedge d\Pi_0,\eeq 
and we take for hamiltonian
\beq\label{gf2}
H=\frac 12\,g^{\mu\nu}\,\Pi_{\mu}\,\Pi_{\nu}=
=\frac 12\left(\frac 1V\,(\Pi_i-\Pi_0\,\T_i)^2+V\,\Pi_0^2\right).\eeq
For geodesics orthogonal to the $U(1)$ fibers and affinely parametrized 
by $\,\la\,$ the equations for 
the flow allow on the one hand to express the velocities
\beq\label{gff3}\barr{l}
\dst\dot{t}\equiv\frac{dt}{d\la}=\frac{\pt H}{\pt \Pi_0}=
\left(V+\frac{\T^2}{V}\right)\Pi_0-\frac{\T_i\Pi_i}{V},\\[4mm]
\dst\dot{x}_i\equiv\frac{dx_i}{d\la}=\frac{\pt H}{\pt \Pi_i}=\frac 1V\,p_i,
\qq\quad p_i=\Pi_i-\Pi_0\,\T_i,
\earr\eeq
and on the other hand to get the dynamical evolution equations
\beq\label{gf3} \barr{l}\dst 
\dot{\Pi}_0=-\frac{\pt H}{\pt t}=0,
\qq\qq\qq q\equiv\Pi_0=\frac{(\dt+\T_i\,\dx_i)}{V},\hfill (a)\\[4mm]\dst 
\dot{\Pi}_i=-\frac{\pt H}{\pt x_i}\qq\quad\Longrightarrow\qq \quad
\dot{p}_i=\left(\frac HV-q^2\right)\,\pt_i V+\frac qV\,(\pt_i\T_s-\pt_s\T_i)\,p_s.
\qq\hfill(b)
\earr\eeq
Relation (\ref{gf3}a) expresses the conservation of the charge 
$\dst\,q,$ a consequence of the $U(1)$ isometry of the metric. 
For the multi-centre metrics, use of relation (\ref{mc3}) brings the 
equations of motion (\ref{gf3}b) to the nice form  
\beq\label{gf6}
\dot{\vec{p}}=\left(\frac HV-q^2\right)\,\vec{\nabla}\,V+
\frac qV\ \vec{p}\wedge\vec{\nabla} V.
\eeq

The conservation of the energy 
\beq\label{gf5}
H=\frac 12\left(\frac{p_i^2}{V}+q^2\,V\right)=\frac V2(\dx_i^2+q^2)=
\frac 12\,g_{\mu\nu}\,\dx^{\mu}\,\dx^{\nu}\eeq
is obvious since it expresses the constancy of the length of the tangent vector 
$\,\dx^{\mu}\,$ along a geodesic.

\subsection{Killing-St\" ackel tensors and their conserved quantities}
A Killing-St\" ackel (KS) tensor is a symmetric tensor $\,S_{\mu\nu}\,$ which 
satisfies
\beq\label{ks1}
\nabla_{(\mu}\,S_{\nu\rho)}=0.\eeq
Let us observe that if $\,K\,$ and $\,L\,$ are two (possibly different) Killing 
vectors their symmetrized tensor product $\,K_{(\mu}\,L_{\nu)}\,$ 
is a KS tensor. So we will define {\em irreducible} KS tensors as the ones which 
cannot be written as linear combinations, with constant coefficients, of 
symmetrized tensor products of Killing vectors. 

For a given KS tensor $\,S_{\mu\nu}\,$ the quadratic form of the velocities:
\beq\label{ks2}
{\cal S}=S_{\mu\nu}\,\dx^{\mu}\,\dx^{\nu}\eeq
is preserved by the geodesic flow. 

In all what follows we will look for KS tensors, under the 
assumptions
\brm
\item[\quad $\mathbf{A 1}$ :] The KS tensor is preserved by Lie dragging 
along the tri-holomorphic Killing vector:
\beq\label{Lie}
\dl_{\ti{\cal K}}\,S_{\mu\nu}=0,\qq\qq \ti{\cal K}=\pt_t\eeq
\item[\quad $\mathbf{A 2}$ :] We will consider {\em generic} values of $\,H\,$ 
and $\,q\neq 0.$  
\erm
Furthermore, instead of focusing ourselves on the KS tensor 
$\,S_{\mu\nu},$ whose usefulness is just to produce the conserved 
quantity ${\cal S},$ let us rather examine more closely the structure of 
the  conserved quantity induced by such a KS tensor.\ From relation 
(\ref{ks2}) we obtain the following ansatz for the conserved quantity 
we are looking for:
\beq\label{ks4}
{\cal S}=A_{ij}(x_k)\,p_i\,p_j+2q\,B_i(x_k)\,p_i+C(x_k),\eeq
where the various unknown functions, as a consequence of A1, are 
independent of the coordinate on the $U(1)$ fiber. 

It is interesting to notice that the knowledge of $\,{\cal S}\,$ is {\em equivalent} 
to the knowledge of the K-S tensor: using (\ref{gff3}) one can express 
$\,{\cal S}\,$ in terms of the velocities and, going backwards, compute the K-S 
tensor components from relation (\ref{ks2}).

\noindent Imposing the conservation of $\,{\cal S}\,$ under the geodesic flow
gives:

\begin{nth}
Under the assumptions A1 and A2 the quantity $\,{\cal S},$ given by
(\ref{ks4}),  is conserved iff the following equations are satisfied
\footnote{The  assumption A2 implies that $\,H-q^2V\,$ does not vanish
identically.} \beq\label{ks}\barr{ll}\dst 
a)\qq & q\cdot \dl_B\,V=0 \hfill \\[4mm]
b)\qq & \pt_{(k}\,A_{ij)}=0\\[4mm]
c)\qq & q(\pt_{(i}\,B_{j)}-A_{s(i}\,\eps_{j)su}\,\pt_uV)=0\\[4mm]
d)\qq & \pt_i\,C+2(H-q^2\,V)\,A_{is}\,\pt_s\,V-2\,q^2\,\eps_{ist}\,B_s\,\pt_tV=0 
\earr\eeq
\end{nth}
We are now in position to explain why we assumed, in A2, that $\,q\,$ should not vanish. 
Indeed for $\,q=0\,$ the relations (\ref{ks}a) and (\ref{ks}c) are trivially true 
and we are left with
\[\pt_{(k}\,A_{ij)}=0,\qq\quad \pt_i\,C+2H\,A_{is}\,\pt_s\,V=0,\] while the conserved 
quantity $\,{\cal S}\,$ reduces to
\[{\cal S}=A_{ij}(x)\,p_i\,p_j+C(x).\]
It is interesting to notice that, formally, $\,{\cal S}\,$ is preserved by the 
hamiltonian flow induced by the classical hamiltonian \cite{Pe}
\[{\cal H}=\frac{\vec{p}\,^2}{2}-H\,V,\]
where now $\,H\,$ appears as some constant parameter. However the assumption that 
$\,q=0\,$ leads to a reduced system which has only three degrees of freedom and as such 
may exhibit integrability. Since we are interested in genuine four dimensional 
integrability we have to exclude such a possibility.

Let us proceed to the discussion of the system (\ref{ks}). 
Relation (\ref{ks}a) shows that there are two possible situations:
\brm
\item Either the potential $V$ has one (or more) spatial symmetries, with Killing 
$\,K,$ and then $\,B\,$ has to be conformal to this Killing vector. 
\item Or the potential has no spatial symmetry, and in this case $\,B=0,$
\erm
Let us show that this last possibility does not give any new conserved 
quantity. Indeed relation (\ref{ks}c) can be written
\beq
[A,R]=0,\qq\qq (R)_{ij}=\eps_{isj}\,\pt_s\,V.\eeq
Since $V$ has no Killing the matrix $R$ is a 
generic matrix in the Lie algebra $so(3).$ By Schur lemma it follows that $A$ 
has to be proportional to the identity matrix and this does trivialize the 
corresponding conserved quantity ${\cal S}.$

So the unique possibility left is the first one. Let us notice that $\,K\,$ 
lifts up to an isometry of the 4 dimensional metric. We have obtained:

\begin{nth}
The number of extra conserved quantities, having the structure (\ref{ks4}), 
of a multi-centre metric is at most equal to the number of extra {\rm spatial} 
Killing vectors it does possess, besides the tri-holomorphic 
Killing $\,\wti{\cal K}=\pt_t$.
\end{nth}
Using this result we can discuss the triaxial generalization of the Eguchi-Hanson 
metric, with a tri-holomorphic $su(2)$, discovered in \cite{bgpp}. Its potential 
and cartesian coordinates were given in \cite{gorv} and the potential has no
spatial Killing. From the previous proposition it  follows that this metric 
will not  exhibit  a conserved quantity of the form (\ref{ks4}) for generic
values of $H$ and $q\neq 0.$

\subsection{Transformations of the system}
As observed above, the vector $\,B\,$ has to be conformal to the Killing $\,K.$ So we 
define the conformal factor $\,F\,$ such that
\beq\label{ks5}
B_i=-\,F\,K_i.\eeq  
The conserved quantity (\ref{ks4}) becomes
\beq\label{cq}
{\cal S}=A_{ij}(x)\,p_i\,p_j-2\,q\,F\,K_i\,p_i+C(x),\eeq
and equation (\ref{ks}c) transforms into
\beq\label{ks6}
K_{(i}\,\pt_{j)}F+A_{s(i}\,\eps_{j)su}\pt_u\,V=0.\eeq
Taking its trace we see that $\,\dst\dl_K\,F=0,$ showing that $\,V\,$ and 
$\,F\,$ must have the same Killing.

Contracting (\ref{ks6}) with $\pt_jV$ gives 
\begin{nlem}\label{l1}
The equation (\ref{ks6}) has for consequence:
\beq\label{ksup}
\quad (dV\cdot dF)K+\star(A[dV]\wedge dV)=0,\qq\qq A[dV]=A_{is}\,\pt_sV\,dx^i.\eeq
\end{nlem}
We can proceed to:
\begin{nth}
The relation (\ref{ks6}) is equivalent (except possibly at the points where 
the norm of the Killing $K$ vanishes) to the relations:
\beq\label{ks7}\left\{\barr{l}
A[K]=a(x)\,K, \hfill a)\\[4mm]
|K|^2\,dF-A[\star(K\wedge dV)]+\star(A[K]\wedge dV)=0\qq\qq\hfill b)
\earr\right.\eeq
\end{nth}

\noindent{\bf Proof :} Contracting relation (\ref{ks6}) with $K_j$ gives 
relation b), while contracting with $K_iK_j$ we have
\beq\label{kss1}
\eps_{stu}K_s A[K]_t \pt_u V=0\qq\Longrightarrow\qq A[K]_i=a(x)K_i+b(x)\pt_i V,\eeq
which is not relation a). To complete the argument we first contract 
relation (\ref{ks6}) with $\eps_{iab}K_a$; after some algebra we get
\beq\label{kss2}
K_j\eps_{iab}\pt_iF K_a+2A[K]_j\pt_b V+A[dV]_b K_j
-K_s A[dV]_s\,\de_{jb}-A_{ss}K_j\pt_bV=0,\eeq
which, upon contraction with $A[K]_b,$ gives eventually
\beq\label{kss3}
(A[K]_s\pt_sV)\,A[K]_i=\{-\eps_{stu}K_s A[K]_t \pt_u F+A_{ss}\,A[K]_t\pt_t V
-A[K]_s A[dV]_s\}\,K_i.\eeq
Let us now suppose that $A[K]_s\pt_s V\neq 0.$ The previous relation shows that in 
(\ref{kss1}) we must have $b(x)=0,$ hence $A[K]_s\pt_s V=0$ which 
is a contradiction. 

\noindent Let us prove that the converse 
is true. From $(\ref{ks7}b)$ we get
\beq\label{kst1}
|K|^2\,K_{(i}\,\pt_{j)}F+(K_{(j}A_{i)s}\,K_t\eps_{tsu}+
A[K]_s\,K_{(j}\eps_{i)su})\pt_u V=0.\eeq
Use of the identity
\beq\label{kst2}
A_{is}K_t\,K_j\eps_{tsu}\pt_u V=(|K|^2\,A_{is}\eps_{jsu}-A[K]_i K_t \eps_{jtu})
\pt_u V\eeq
and of relation $(\ref{ks7}a)$ leaves us with $(\ref{ks6}),$ up to division by   
$\,|K|^2.$ Notice that $\,|K|^2\,$ vanishes at the fixed points under the 
Killing action, i. e. in subsets of zero measure in ${\mb R}^3.$ \black

\noindent We can give, using (\ref{ks7}a) and the identity
\beq\label{id1}
-A[\star(K\wedge dV)]=\star(A[K]\wedge dV)-A_{ss}\,\star(K\wedge dV)
+\star(K\wedge A[dV]),\eeq
a simpler form to the relation (\ref{ks7}b):

\begin{nlem}
The relation (\ref{ks7}b) is equivalent to
\beq\label{ksup3}
|K|^2\,dF+(2a-{\rm Tr}\,A)\,\star(K\wedge dV)+\star(K\wedge A[dV])=0.\eeq
\end{nlem}
For further use let us prove:

\begin{nlem}\label{l2} 
To the spatial Killing $\,K,$ leaving the potential $V$ invariant, there 
corresponds a quantity $\,Q\,$ invariant under the geodesic flow given by
\beq\label{sup1}
Q=K_i\,p_i+qG,\qq\mbox{with}\qq i(K)F=-\,dG.\eeq
\end{nlem}

{\bf Proof :} We start from $\,\dst\dl_K\,V=0.$ Since $K$ is a Killing we have 
$\,\dst\dl_K(\star\,dV)=\star\,d(\dl_K\,V)=0,$ and (\ref{mc3}) implies that  
$\,\dst\dl_K\,d\T=0.$ The closedness of $\,d\T\,$ implies $\,d(i(K)d\T)=0,$ and 
since our analysis is purely local in $\,{\mb R}^3,$ we can define 
\beq\label{ksup2}
\,\eta\,dG=-i(K)\,d\T,\qq \Longrightarrow \qq \star(K\wedge dV)=dG.\eeq
Then we multiply (\ref{gf3}b) by $\,p_i$ and get successively
\[K_i\dep_i=\dot{(K_ip_i)}-\dot{K}_i\,p_i=\dot{(K_ip_i)}=
\frac qV K_i\,(\pt_i\T_s-\pt_s\T_i)p_s
=-q\,\dx_s\pt_s G=- q\,\dot{G},\]
which concludes the proof. \black

\noindent Let us point out that if we use the coordinate $\phi$ adapted to 
the Killing $\tilde{K}=\pt_{\phi},$ we can write the connection $\,\T=G\,d\phi,$
 where $\,G\,$ does not depend on $\,\phi.$

\subsection{Integrability equations}
We will derive now the integrability conditions for the equations (\ref{ks}c) and 
(\ref{ks}d). The first one was written using forms in (\ref{ksup3}) while the 
second one is
\beq\label{extra1}
dC+2(H-q^2 V)A[dV]+2q^2 F\,\star(K\wedge dV)=0.\eeq
It can now be proved :

\begin{nth} The integrability condition for (\ref{extra1}) is 
\beq\label{ks8}
d\,A[dV]=0\qq\qq\Longrightarrow\qq A[dV]=dU\qq\mbox{and}\qq \dl_K U=0.\eeq 
\end{nth}

{\bf Proof :} The integrability condition is obtained by 
differentiating (\ref{extra1}). We get
\beq\label{ks9}
2(H-q^2V)\,d\,A[dV]+2q^2\,A[dV]\wedge dV+2q^2\,dF\wedge\star(K\wedge dV)
+ 2q^2F\,d\star(K\wedge dV)=0.\eeq
The last term in this equation vanishes in view of (\ref{ksup2}). Furthermore 
we have the identity specific to three dimensional spaces
\[dF\wedge\star(K\wedge dV)=-(K\cdot dF)\,\star dV+(dV\cdot dF)\,\star K
=(dV\cdot dF)\,\star K\]
because $K$ is a symmetry of $F.$ Relation (\ref{ks9}) simplifies to
\[2(H-q^2 V)\,d\,A[dV]+2q^2\,\star[(dV\cdot dF)\,K+\star(A[dV]\wedge dV)]=0,\]
and lemma \ref{l1} implies the closedness of $A[dV].$ 
Since our analysis is purely local, the existence of $U$ is a consequence of  
Poincar\'e's lemma. 

The relations 
\[\dl_K U=i(K)\,dU=i(K)\,A[dV]=(A[K]\cdot dV)=a(K\cdot dV)=a\,\dl_K V=0\]
show the invariance of $U$ under the Killing $K.$ \black

Let us now turn to equation (\ref{ksup3}). We will prove:

\begin{nth}\label{prop1}
The integrability condition for (\ref{ksup3}) is
\beq\label{ks10}
(2a-{\rm Tr}\,A)dV+dU=|K|^2\,\star d\tau,\qq\qq \dl_K\,d\tau=0,\eeq
for some one form $\tau.$
\end{nth}

{\bf Proof :} Let us define the 1-form
\beq\label{pr1}
Y=(2a-{\rm Tr}\,A)dV+dU.\eeq
It allows to write (\ref{ksup3}) and its integrability condition as
\beq\label{pr2}
dF=-\star\left(\frac{K\wedge Y}{|K|^2}\right),\qq\qq 
\de\left(\frac{K\wedge Y}{|K|^2}\right)=0,\eeq
or switching to components
\beq\label{pr4}
K_i\,\de\left(\frac{Y}{|K|^2}\right)+\frac{Y_s\pt_s K_i-K_s\pt_sY_i}{|K|^2}=0.\eeq
Let us examine the last terms. Since $a$ and ${\rm Tr}\,A$ are invariant under the 
Killing $K,$ we obtain
\beq\label{pr5}
Y_s\pt_sK_i-K_s\pt_sY_i=-(2a-{\rm Tr}\,A)\pt_i(K_s\pt_s\,V)
-\pt_i(K_s\pt_s\,U)\eeq
and both terms vanish because $V$ and $U$ are invariant under $K.$ We are left with 
the vanishing of the divergence of $Y/|K|^2$ from which we conclude (local analysis!) 
that it must have the structure $\,\star d\tau$ for some 1-form $\tau.$\ >From its 
definition it follows that $\,d\tau$ is invariant under $\,K.$ \black

Using this result we can simplify (\ref{ksup3}) to 
\beq\label{ks11}
dF+\star(K\wedge\star d\tau)=dF-i(K)d\tau=0.\eeq
Collecting all these results we have:
 
\begin{nth}
Under the assumptions A1 and A2, the quantity
\[{\cal S}=A_{ij}(x)\,p_i\,p_j-2\,q\,F\,K_i\,p_i+C(x)\]
is preserved by the geodesic flow of the multi-centre metrics provided that 
the integrability constraints
\beq\label{ksf2}
\De\,V=0,\qq A[dV]=dU,\qq (2a-{\rm Tr}\,A)\,dV+dU=|K|^2\,\star d\tau\eeq
and the following relations hold:
\beq\label{ksf1}\barr{ll}\dst 
a)\qq & \dst \dl_K V=0,\\[4mm]
b)\qq & \pt_{(k}A_{ij)}=0,\qq\qq  A[K]=a\,K, \\[4mm]
c)\qq & dF=i(K)\,d\tau,\\[4mm] 
d)\qq & d(C+2HU)+2q^2(-V\,dU+F\,dG)=0,\qq\quad \star(K\wedge dV)=dG.
\earr\eeq
\end{nth}

\subsection{Classification of the spatial Killing vectors}
An important point, in view of classification, is 
whether the extra spatial Killing is tri-holomorphic 
or not. This can be checked thanks to:
\begin{nlem}\label{l3}
The spatial Killing vector $\,K_i\pt_i\,$ is tri-holomorphic iff 
\[\eps_{ist}\pt_{[s}K_{t]}=0.\]
Otherwise it is holomorphic.
\end{nlem}

{\bf Proof :}\ From \cite{bf} we know that, for an hyperk\" ahler geometry, a 
Killing may be either holomorphic or tri-holomorphic. As shown in \cite{gr} 
such a vector will be tri-holomorphic iff the differential of the dual 1-form 
$\,K=K_i\,dx_i\,$ has the self-duality opposite to that of the complex 
structures. A computation shows that this is equivalent to the vanishing of
\[dK^{(-)}=-\frac 1{2}\,\eps_{ijk}\,\pt_{[j}\,K_{k]}
\left(E_0\wedge E_i-\frac 1{2}\,\eps_{ist}\,E_s\wedge E_t\right),\]
from which the lemma follows. \black

Since we are working in a flat three dimensional flat space, there are 
essentially two different cases to consider:
\brm
\item The Killing $K$ generates a spatial rotation, which we can take, without 
loss of generality, around the z axis. In this case we have
\[K_i\,p_i=L_z\]
and this Killing vector is holomorphic with respect to the complex structure 
$\,J_3,$ defined in section 2.
\item The Killing $K$ generates a spatial translation, which we can take, without 
loss of generality, along the z axis. In this case we have the 
\[K_i\,p_i=p_z\]
and this Killing vector is tri-holomorphic.
\erm
We will discuss successively these two possibilities, under the simplifying 
additional assumption:
\brm
\item[$\mathbf{A 3}$ :] the K-S tensor $\,S_{\mu\nu}\,$ is also preserved by Lie 
dragging along the extra spatial Killing vector $K$
\[\dl_{K}\,S_{\mu\nu}=0.\]
\erm

\section{One extra holomorphic spatial Killing vector}
The equation (\ref{ksf1}b) states that $\,A_{ij}\,$ is a Killing tensor in flat 
space. As shown in \cite{kl} such a Killing tensor is totally reducible to 
symmetrized tensor products of Killing vectors and involves 20 
free parameters. It is most conveniently written in terms of 
$\,{\cal A}(p,p)\equiv A_{ij}\,p^i\,p^j.$ One has:
\beq\label{rk1}
{\cal A}(p,p)=\left\{\barr{l}
\alf\,L_x^2+\be\,L_y^2+\ga\,L_z^2+2\mu\,L_y L_z +2\nu\,L_z L_x+2\la\,L_x L_y\\[4mm]
+a_1\,p_x L_y+a_2\,p_x L_z+b_1\,p_y L_x+b_2\,p_y L_z+c_1\,p_z L_x+c_2\,p_z L_y\\[4mm]
+d_1\,p_x L_x+d_2\,p_y L_y+a_{ij}p_ip_j.\earr\right.\eeq
The constraint (A 3) for the rotational Killing, requires $\,\dst\dl_K\,A_{ij}=0,$ 
which allows to bring (\ref{rk1}) to the form
\beq\label{rk2}
{\cal A}(p,p)=\alf(L_x^2+L_y^2)+\ga\,L_z^2
+b\,(\vec{p}\wedge \vec{L})_z+a_{33}\,p_z^2+a_{11}\vec{p}\,^2+\de\,p_z L_z.\eeq
We note that the parameter $\,\ga\,$ corresponds to a {\em reducible} piece which 
is just the square of $\,L_z.$ We will take $\,\ga=\alf\,$ for convenience. 

The parameter $\,a_{11}\,$ is easily seen, upon integration 
of the remaining equations in (\ref{ks}), to give rise, in the conserved 
quantity $\,{\cal S},$ to the full piece
\beq\label{rk3}
a_{11}(\vec{p}\,^2-2HV+q^2V^2)\eeq
which vanishes thanks to the energy conservation (\ref{gf5}). So we can take 
$\,a_{11}=0.$ 

The second relation in (\ref{ksf1}b) implies the vanishing of $\,\de.$ Hence, 
with slight changes in the notation, we end up with \beq\label{rk6}
{\cal A}(p,p)=a\,\vec{L}\,^2+c^2\,p_z^2+b\,(\vec{p}\wedge\vec{L})_z.\eeq
Let us note that the parameters $a$ and $b$ are real while the parameter $c$ 
may be either real or pure imaginary.

To take advantage of the rotational symmetry around the z axis we use the 
coordinates 
\[x=\sqrt{\rho}\,\cos\phi,\qq y=\sqrt{\rho}\,\sin\phi,\qq z,\]
and write the connection
\[\T=G\,d\phi.\]
By lemma 3, this symmetry gives for conserved quantity
\beq\label{cq1}
\,J_z=L_z+q\,G=x\,\Pi_y-y\,\Pi_x.\eeq 

>From the system (\ref{ksf1}) one can 
check that the functions $F$ and $U$ are to be determined from
\beq\label{fc1a}\left\{\barr{l}
F_{,\rho}=(az+b/2)V_{,z}-a/2\,V_{,z}\\[4mm] 
F_{,z}=2(az^2+bz-c^2)V_{,\rho}-(az+b/2)V_{,z}\earr\right.\eeq
and 
\beq\label{fc1b}\left\{\barr{l}
U_{,\rho}=z(az+b)V_{,\rho}-\frac 12(az+b/2)V_{,z}\\[4mm]
U_{,z}=-2\rho(az+b/2)V_{,\rho}+(a\rho+c^2)V_{,z}\earr\right.\eeq

\subsection{The two-centre metric}
This case corresponds to the choice $\,a=1\,$ and $\,c\neq 0.$ 
Since $a=1,$ we can get rid of the constant $b$ by a 
translation of the variable $z.$ So, without loss of generality, we can take 
$b=0$ and use the new variables 
$\,r_{\pm}=\sqrt{x^2+y^2+(z\pm c)^2}.$ We get the relations
\[\pt_{r_+}F=-c\,\pt_{r_+} V,\qq\quad \pt_{r_-}F=+c\,\pt_{r_-} V\]
which imply
\[V=f(r_+)+g(r_-),\qq\qq F=-c(f(r_+)-g(r_-)).\]
Imposing to the potential $V$ the Laplace equation we have
\beq\label{fc2}
V=v_0+\frac{m_1}{r_+}+\frac{m_2}{r_-},
\qq\qq F=-c\left(\frac{m_1}{r_+}-\frac{m_2}{r_-}\right)=-c\De,\eeq
i. e. we recover the most general 2-centre metric. Let us recall that only the 
double Taub-NUT metric, given by real $\,m_1=m_2,$ is complete. If in addition 
we take the limit $\,v_0\to 0,$ we are led to the Eguchi-Hanson \cite{eh} metric.

One has then to check the integrability constraint (\ref{ks8}) and to determine 
the functions $U$ and $C$ \footnote{We discard constant 
terms in the function $C.$}
\beq\label{fc3}
U=-cz\De,\qq\qq
C=-2(H-q^2\,V)U-q^2\,r^2\,\De^2,\qq r^2=x^2+y^2+z^2.\eeq
Let us observe that the conserved quantity which we obtain may be real even if 
$c$ is pure imaginary. In this case $m_1=m$ may be complex, but if we take 
$m_2=m^{\star}$ the functions $V$ and $c\De$ are real, as well as ${\cal S}.$ One 
obtains quite different metrics (as first observed in the particular case 
of Eguchi-Hanson metric): real $c$ corresponding to type II metric and
$c$ pure  imaginary for type I metric, in the terminology of \cite{eh}.

The final form of the conserved quantity for the two-centre metric is therefore
\beq\label{fc4}\left\{\barr{l}\dst 
{\cal S}_I=\vec{L}\,^2+c^2\,p_z^2+2\,qc\,\De\,L_z+2cz\,\De\,(H-q^2V)
-q^2 r^2\De^2, \\[4mm]
\dst V=v_0+\frac{m_1}{r_+}+\frac{m_2}{r_-}, \qq  \De=\frac{m_1}{r_+}-\frac{m_2}{r_-}, 
\qq G=m_1\,\frac{z+c}{r_+}+m_2\,\frac{z-c}{r_-}.\earr\right.\eeq

The relation of our results with the separability of the Hamilton-Jacobi equation for 
the two-centre metric, obtained in \cite{gr}, will be discussed in the next section.

\ From the very definition of the coordinates $r_{\pm}$ it is clear that 
the previous analysis is only valid for $c\neq 0.$ The 
special case $c=0$ (it is a {\em singular} limit), giving a first dipolar  
breaking of the Taub-NUT metric, will be examined now. 

\subsection{First dipolar breaking of Taub-NUT}
This case corresponds to the choice $\,a=1\,$ and $\,c=0.$ 
Since $a=1,$ we can again get rid of the parameter $b.$ Then 
relation (\ref{fc1a}) for $F$ implies 
\beq\label{sc1}
V=w_0(r)+w_1(r)\,z,\qq\qq F_{,r}=-rw_1(r),\qq r=\sqrt{x^2+y^2+z^2}.\eeq
Imposing the Laplace equation we obtain
\beq\label{sc2}
V=v_0+\frac mr+{\cal E}z+{\cal F}\frac z{r^3},\qq\qq 
F=-\frac{\cal E}{2}\,r^2+\frac{\cal F}{r}.\eeq
The integrability relations for $U$ require that ${\cal E}=0$ and we have
\beq\label{sc3}
U={\cal F}\frac zr,\qq\qq C=-2{\cal F}\,\frac zr\,(H-q^2V)
-2mq^2{\cal F}\,\frac z{r^2}-q^2{\cal F}^2\,\frac{(3z^2-r^2)}{r^4}.\eeq

The final form of the conserved quantity is therefore
\beq\label{sc4}\left\{\barr{l}\dst
{\cal S}_{II}=\vec{L}\,^2-2\, q\,\frac{\cal F}{r}\,L_z
-2{\cal F}\,\frac zr\,(H-q^2v_0)+q^2{\cal F}^2\frac{(x^2+y^2)}{r^4},\\[4mm]
\dst V=v_0+\frac mr+{\cal F}\frac z{r^3},\qq 
G=m\frac zr-{\cal F}\,\frac{x^2+y^2}{r^3}.\earr\right.\eeq
Let us now consider the case $\,a=0,$ which leads to a second dipolar breaking 
of Taub-NUT.

\subsection{Second dipolar breaking of Taub-NUT}
This case corresponds to the choice $\,a=0\,$ and $\,b=1.$ 
The relation (\ref{fc1a}) shows that by a translation of $z$ we can take, without 
loss of generality, $c=0.$ From the integrability of $F$ we deduce
\[
V=\wh{f}(x^2+y^2+(z-c)^2)+g(z). \]
Hence by a translation of $z$ we can set $c$ to 0. We are left with
\beq\label{tc1}
V=f(r)+g(z),\qq\qq F=\frac 12(f(r)-g(z)).\eeq
Imposing Laplace equation yields
\beq\label{tc2}
V=v_0+\frac mr+{\cal E}z,\qq\qq F=\frac 12\left(\frac mr-{\cal E}z\right)\eeq
Then the integrability conditions for $U$ and $C$ are satisfied and we obtain
\beq\label{tc3}
U=\frac{mz}{2r}-\frac{\cal E}{4}\,(x^2+y^2),\qq C=-2U(H-q^2 v_0)
-2q^2\,m{\cal E}\,\frac{(x^2+y^2)}{r}.\eeq

The final form of the conserved quantity is therefore
\beq\label{tc4}\left\{\barr{l}\dst
{\cal S}_{III}=(\vec{p}\wedge\vec{L})_z
-q\,\left(\frac mr-{\cal E}\,z\right)L_z
-2U\,(H-q^2 v_0)-2q^2\,m{\cal E}\,\frac{(x^2+y^2)}{r} \\[4mm]\dst 
V=v_0+\frac mr+{\cal E}\,z\qq\qq G=m\frac zr+\frac{\cal E}{2}(x^2+y^2).
\earr\right.\eeq

For ${\cal E}=0$ we are back to the Taub-NUT metric. In this case the spatial 
isometries are lifted up from $u(1)$ to $su(2).$ As a result we have now 
three possible Killings to start with
\beq\label{ko}
K_i^{(1)}p_i=L_x\qq K_i^{(2)}p_i=L_y\qq K_i^{(3)}p_i=L_z\eeq
and we expect that the conserved quantity found above should be part of a triplet. 
The two missing conserved quantities can be constructed following the same 
route which led to ${\cal S}_{III}$ using the new available spatial 
Killings given by (\ref{ko}). We recover \beq\label{rg1}
\vec{\cal S}=\vec{p}\wedge\vec{L}-q\,\frac mr\,\vec{L}
+m(q^2v_0-H)\frac{\vec{r}}{r},\qq{\cal S}_{III}({\cal E}=0)\equiv{\cal S}_z.\eeq
Lemma \ref{l2} lifts up $\,J_z,$ given by (\ref{cq1}), to a triplet 
of conserved quantities
\beq\label{rg2}
\vec{J}=\vec{L}+q\,\frac mr\,\vec{r},\eeq
which allows to write
\beq\label{rg3}
\vec{\cal S}=\vec{p}\wedge\vec{J}+m(q^2v_0-H)\frac{\vec{r}}{r},\eeq
on which we recognize the generalized Runge-Lenz vector discovered by 
Gibbons and Manton \cite{gm}.

We have therefore obtained, for the three hamiltonians 
$H_I,\,H_{II}({\cal F}\neq 0)\,$ and $\,H_{III},$ corresponding respectively 
to the extra conserved quantities $\,{\cal S}_I,\,{\cal S}_{II}$ and 
$\,{\cal S}_{III},$  (the proof of their irreducibility with respect to the Killing 
vectors is easy) a set of four independent conserved quantities:
\[H,\quad q=\Pi_0,\quad J_z,\quad{\cal S},\]
which can be checked to be in involution with respect to the Poisson bracket. 

Hence we conclude to:

\begin{nth}
The three hamiltonians $H_I,\,H_{II}({\cal F}\neq 0)\,$ and $\,H_{III},$  defined 
above are integrable in Liouville sense.
\end{nth}

\section{One extra tri-holomorphic spatial Killing vector}
This time we have for Killing $K_ip_i=p_z.$ Imposing (A 3) for the translational 
invariance and the constraint $A[K]\propto K$ restricts  
${\cal A}(p,p)$ to have the form
\beq\label{tk1}  
{\cal A}(p,p)=a\,L_z^2-2b\,p_xL_z+2c\,p_yL_z+\sum_{i,j=1}^2\,a_{ij}\,p_ip_j.\eeq
We have omitted a term proportional to $p_z^2$ since it is reducible.

The functions $\,F\,$ and $\,U,$ which depend only on the coordinates $x$ and $y,$ 
using the system (\ref{ksf1}), are seen to be determined by
\beq\label{th1}
\left\{\barr{l}
F_{,x}=A_{12}\,V_{,x}-A_{11}\,V_{,y}\\[4mm]
F_{,y}=A_{22}\,V_{,x}-A_{12}\,V_{,y}\earr\right.\qq
\left\{\barr{l}
U_{,x}=A_{11}\,V_{,x}+A_{12}\,V_{,y}\\[4mm]
U_{,y}=A_{12}\,V_{,x}+A_{22}\,V_{,y}\earr\right.\eeq
with
\beq\label{th2}
A_{11}=ay^2+2by+a_{11},\qq A_{22}=ax^2+2cx+a_{22},\qq A_{12}=-axy-bx-cy+a_{12}.\eeq
\noindent In order to organize the subsequent discussion, let us observe:
\brm
\item For $\,a\neq 0,$ we may take $\,a=1.$ The spatial translations in the xy-plane 
allow to take $\,b=c=0,$  and a rotation  $a_{12}=0$ as well. Hence we are left with
\[{\cal A}(p,p)=L_z^2+(a_{11}-a_{22})\,p_x^2+a_{22}(p_x^2+p_y^2).\]
Adding the reducible term $\,a_{22}\,p_z^2$ we recover the piece 
$a_{22}\,\vec{p}\,^2\,$ which can be discarded, as already explained in section 4. 
So we will take for our first case
\beq\label{thk1}
{\cal A}_1(p,p)=L_z^2-c^2\,p_x^2,\qq c\in{\mb R}\cup i{\mb R}, \qq c\neq 0.\eeq
\item Our second case, which is the singular limit $\,c\to 0\,$ of the first case, 
corresponds to
\beq\label{thk2}
{\cal A}_2(p,p)=L_z^2.\eeq
\item For $\,a=0,$ a first translation allows to take $\,a_{12}=0,$ while the 
second one allows the choice $\,a_{11}=a_{22}\,$ and the corresponding term 
$\,a_{11}(p_x^2+p_y^2)\,$ is disposed of as in the first case. Eventually a rotation 
will bring $b$ to zero and $c$ to $1.$ Our third case will be
\beq\label{thk3}
{\cal A}_3(p,p)=p_y\,L_z.\eeq
\item For $a=b=c=0,$ we can discard $p_x^2+p_y^2$  and we are left with 
our fourth case
\beq\label{thk4}
{\cal A}_4(p,p)=\alf\,p_y^2+\be\,p_x\,p_y.\eeq
\erm

We will state the results obtained for these four cases without going through 
the detailed computations, which are greatly simplified by the use of the complex 
coordinate $w=x+iy.$ In all four cases the metric will have the form
\beq\label{thk5}
g=\frac 1V(dt+\T)^2+V(dz^2+d\ol{w}dw),\qq \T=G\,dz.\eeq

\subsection{First case}
Writing the conserved quantity as
\beq\label{Fc1}
{\cal S}_1=L_z^2-c^2\,\Pi_x^2-2c^2 F\,\Pi_0\,\Pi_z+c^2(2v_0U+D)\Pi_0^2-2c^2UH ,
\qq c\neq 0,\eeq
where $\,\Pi_z=p_z+G\,\Pi_0\,$ and
\beq\label{Fc2}\barr{l}\dst
\bullet\quad V+iG=v_0+2m\,\frac{w}{\sqrt{w^2+c^2}},\qq v_0\in{\mb R},
\quad m\in{\mb C}\\[4mm]\dst
\bullet\quad U+iF=-m\,\frac{w+\ol{w}}{\sqrt{w^2+c^2}},\qq 
D=-2|m|^2\,\frac{(w^2+\ol{w}^2+|w|^2+c^2)}{|\sqrt{w^2+c^2}|^2}.\earr\eeq

\subsection{Second case}
Writing the conserved quantity as
\beq\label{Sc1}
{\cal S}_2=L_z^2-2F\,\Pi_0\,\Pi_z+2v_0U\,\Pi_0^2-2UH,\eeq
we have:
\beq\label{Sc2}\barr{l}\dst
\bullet\quad V+iG=v_0+\frac{m}{w^2},
\qq v_0\in{\mb R},\quad m\in{\mb C},\\[4mm]\dst
\bullet\quad U+iF=m\,\frac{\ol{w}}{w}.
\earr\eeq

\subsection{Third case}
Writing the conserved quantity as
\beq\label{Tc1}
{\cal S}_3=\Pi_y\,L_z-2F\,\Pi_0\,\Pi_z+(2v_0U+D)\Pi_0^2-2UH,\eeq
we have:
\beq\label{Tc2}\barr{l}\dst
\bullet\qq V+iG=v_0+2\,\frac{m}{\sqrt{w}},
\qq v_0\in{\mb R},\quad m\in{\mb C},\\[4mm]\dst
\bullet\quad U+iF=-\frac m2\,\frac{w-\ol{w}}{\sqrt{w}},\qq 
D=|m|^2\,\frac{w+\ol{w}}{|\sqrt{w}|^2}.\earr\eeq

\subsection{Fourth case}
In this case we take for the driving term 
\[{\cal A}_4(p,p)= \alf\,p_y^2+\be\,p_x\,p_y.\]
Using the freedom of rotations in the xy-plane, at the level of the metric, we 
can take
\beq\label{pt4}
V=v_0+mx,\qq\qq G=my.\eeq
This time there are {\em two} conserved quantities
\beq\label{Ffc1}
{\cal S}_4=\alf\,{\cal S}^{(1)}_4+\be\,{\cal S}^{(2)}_4,\eeq 
given by
\beq\label{Ffc2}\left\{\barr{l}
{\cal S}^{(1)}_4=\Pi_y^2+(\Pi_z-my\,\Pi_0)^2,\\[4mm]
{\cal S}^{(2)}_4=\Pi_x\,\Pi_y-V\,\Pi_0(\Pi_z-my\,\Pi_0)-my\,H.\earr\right.\eeq
We added reducible terms of the form $\Pi_z^2$ and $\Pi_z\,\Pi_0$  
to get a simpler final form.

The metric exhibits one further tri-holomorphic Killing vector and a 
corresponding conserved quantity  
\[\pt_y- mz\,\pt_t\qq\Rightarrow\qq \Pi_y-mz\,\Pi_0.\]

\noindent Let us close the algebra of the conserved quantities under 
Poisson bracket. For the Killing vectors we recover a Bianchi II Lie algebra
\[\{\Pi_0,\Pi_z\}=0,\qq\quad\{\Pi_z,\Pi_y-mz\,\Pi_0\}= m\,\Pi_0,
\qq\quad\{\Pi_y-mz\,\Pi_0,\Pi_0\}=0.\]
The K-S tensors are invariant under the Killing vectors action, and it may be 
interesting to note that the Schouten bracket of the two K-S tensors is vanishing. 
This hamiltonian is therefore super-integrable. 

To conclude this section let us notice that, among the four potentials considered, 
only the second one and the fourth one are {\em uniform} functions in the 
three dimensional flat space.

As was the case when the extra spatial Killing was holomorphic, we have obtained 
for the four hamiltonians considered in this section, a set of (at least) 
four conserved quantities
\[H,\quad q=\Pi_0,\quad \Pi_z,\quad {\cal S},\]
and in all the four cases $\,{\cal S}\,$ is irreducible with respect to the 
Killing vectors. One can check that these four independent conserved quantities 
are in involution with respect to the Poisson bracket, hence we have:

\begin{nth}
The four hamiltonians determined in this section are integrable in Liouville sense.
\end{nth}

As is well known the existence of K-S tensors is related to the separability of the 
Hamilton-Jacobi (H-J) equation, or equivalently to the separability of the 
Schr\" odinger equation. In the next sections we will analyze the separability of 
the H-J equation according to the nature of the extra Killing vector.

\section{H-J separability: extra holomorphic Killing}
We write the metric
\beq\label{met1}
g=\frac 1V(dt+G\,d\phi)^2+V(\ga_1\,d\xi_1^2+\ga_2\,d\xi_2^2+\ga_3\,d\phi^2),\eeq
which makes apparent the two commuting Killing vectors $\,\wti{\cal K}=\pt_t\,$ and 
$\,\wti{\cal L}=\pt_{\phi},$ where only the first one is tri-holomorphic.

The hamiltonian is
\beq\label{ham1}
H=\frac{G^2+\ga_3\,V^2}{2\ga_3V}\,\Pi_0^2-\frac{G}{\ga_3V}\,\Pi_0\,\Pi_{\phi}
+\frac{\Pi_{\phi}^2}{2\ga_3V}
+\frac 1{2V}\left(\frac{\Pi_1^2}{\ga_1}+\frac{\Pi_2^2}{\ga_2}\right).\eeq
Since the $\,\ga_i$'s depend only on $\xi_1$ and $\xi_2,$ it follows that 
$\,\Pi_0\,$ and $\,\Pi_{\phi}\,$ are conserved.

\subsection{The two-centre case}
The H-J equation separability was first used in \cite{gr} to get the corresponding 
K-S tensor. This reference is muddied by so many misprints that we will 
present its results anew.  

Separability relies here on the use of spheroidal coordinates 
$\xi_1=\z,\,\xi_2=\la,$ defined by
\[x= c\sqrt{(\z^2-1)(1-\la^2)}\,\cos\phi,\qq
y=c\sqrt{(\z^2-1)(1-\la^2)}\,\sin\phi,\qq z=c\,\z\la.\]
This implies
\[\ga_1=c^2\frac{\z^2-\la^2}{\z^2-1},\qq\ga_2=c^2\frac{\z^2-\la^2}{1-\la^2},\qq
\ga_3=c^2(\z^2-1)(1-\la^2).\]

The potential and connection are
\beq\label{potconn}
V=v_0+\frac{\si\z-\de\la}{c(\z^2-\la^2)},\qq
G=\frac{\si\la(\z^2-1)+\de\z(1-\la^2)}{\z^2-\la^2},\eeq
with $\,\si=m_1+m_2\,$ and $\,\de=m_1-m_2.$ 

The hamiltonian is
\beq\label{tchj1}
H=\frac 1{2c^2\,V}\left\{\frac{(\z^2-1)\,\Pi_{\z}^2
+(1-\la^2)\,\Pi_{\la}^2}{(\z^2-\la^2)}
+\frac{(\Pi_{\phi}-G\,\Pi_0)^2}{(\z^2-1)(1-\la^2)}\right\}+\frac V2\,\Pi_0^2.\eeq
The separation constants  \footnote{In all what follows each couple of 
separation constants add up to zero} are
\beq\label{tchj2}\barr{l}\dst 
C_{\z}=(\z^2-1)\,\Pi_{\z}^2+\frac{\Pi_{\phi}^2}{\z^2-1}
-2\,\de\frac{\z}{\z^2-1}\,\Pi_0\,\Pi_{\phi}-2c(v_0c\z^2+\si\z)H\\[4mm]\dst 
\hspace{7cm} +\left(\frac{\de^2}{\z^2-1}+v_0^2c^2\z^2+2v_0c\si\z\right)\Pi_0^2,
\earr\eeq
and
\beq\label{tchj3}\barr{l}\dst 
C_{\la}=(1-\la^2)\,\Pi_{\la}^2+\frac{\Pi_{\phi}^2}{1-\la^2}
-2\,\si \frac{\la}{1-\la^2}\,\Pi_0\,\Pi_{\phi}+2c(v_0c\la^2+\de\la)H\\[4mm]\dst 
\hspace{7cm}+\left(\frac{\si^2}{1-\la^2}
-v_0^2c^2\la^2-2v_0c\de\la\right)\Pi_0^2 .\earr\eeq

The knowledge of these separation constants is of paramount importance since 
it reduces the integration of the H-J equation to quadratures. Indeed writing
\[S=t\,\Pi_0+\phi\,\Pi_{\phi}+A(\z)+B(\la),\]
one has just to replace $\,\Pi_{\z}\,$ by 
$\,\dst\frac{dA}{d\z}\,$ in (\ref{tchj2}) and $\,\Pi_{\la}\,$ by 
$\,\dst\frac{dB}{d\la}\,$ in (\ref{tchj3}) to get the relevant separated differential 
equations. In practice the final integrations may be quite tough.

Some algebra allows to relate the conserved quantity obtained in section 3  
to the separation constants, with the final simple result
\beq\label{tchj4}
{\cal S}_I=C_{\la}-(\si^2+\de^2)\Pi_0^2.\eeq
In \cite{gr} it was conjectured that in the Taub-NUT limit $\,c\to 0\,$ this 
separation constant could be related to some component of the 
generalized Runge-Lenz vector. We can check that this is not true since, using 
relation (\ref{fc4}), we get
\beq\label{tchj5}
\lim_{c\to 0}\ {\cal S}_I=\vec{L}\,^2-\de^2\,\Pi_0^2.\eeq

\subsection{First dipolar breaking}
The H-J equation does separate in spherical coordinates $\,\xi_1=r,\,\xi_2=\tht,$ 
for which we have
\[\ga_1=1,\qq \ga_2=r^2,\qq \ga_3=r^2\sin^2\tht,\]
and
\beq\label{fdb1}
V=v_0+\frac mr+{\cal F}\,\frac{\cos\tht}{r^2},\qq 
G=m\,\cos\tht-{\cal F}\,\frac{\sin^2\tht}{r}.\eeq
The separation constants in the H-J equation are
\beq\label{fdb3}
C_r=r^2\,\Pi_r^2+2\,\frac{\cal F}{r}\,\Pi_0\,\Pi_{\phi}
+\left(v_0^2r^2+2v_0mr+\frac{{\cal F}^2}{r^2}\right)\Pi_0^2-2(v_0r^2+mr)H,\eeq 
and
\beq\label{fdb4}
C_{\tht}=\Pi_{\tht}^2+\frac{\Pi_{\phi}^2}{\sin^2\tht}
-2\, m\,\frac{\cos\tht}{\sin^2\tht}\,\Pi_0\,\Pi_{\phi}+\left(\frac{m^2}{\sin^2\tht}
+2v_0{\cal F}\cos\tht\right)\Pi_0^2-2{\cal F}\,\cos\tht\,H.\eeq
The relation with the K-S tensor of section 3 is $\ 
{\cal S}_{II}=C_{\tht}-m^2\,\Pi_0^2.$

\subsection{Second dipolar breaking}
The H-J equation does separate in parabolic coordinates $\,\xi_1=\xi,\,\xi_2=\eta,$ 
for which we have
\[\ga_1=\frac{(\xi+\eta)}{4\xi},\qq \ga_2=\frac{(\xi+\eta)}{4\eta},\qq 
\ga_3=\xi\,\eta,\]
and
\beq\label{sdb1}
V=v_0+\frac{2m}{\xi+\eta}+\frac{\cal E}{2}(\xi-\eta),\qq 
G=m\,\frac{\xi-\eta}{\xi+\eta}+\frac{\cal E}{2}\,\xi\,\eta.\eeq
The separation constants in the H-J equation are
\beq\label{sdb2}\barr{l}\dst 
C_{\xi}=4\xi\,\Pi_{\xi}^2 +\frac{\Pi_{\phi}^2}{\xi}
+2\left(\frac m{\xi}-\frac{\cal E}{2}\,\xi\right)\,\Pi_0\,\Pi_{\phi}
-2\left(m+v_0\xi+\frac{\cal E}{2}\xi^2\right)H\\[4mm]\dst \hspace{4cm}
+\left(\frac{m^2}{\xi}+2v_0m+(v_0^2+3m{\cal E})\xi+v_0{\cal E}\xi^2
+\frac{{\cal E}^2}{4}\xi^3\right)\Pi_0^2,\earr\eeq
and
\beq\label{sdb3}\barr{l}\dst 
C_{\eta}=4\eta\,\Pi_{\eta}^2+\frac{\Pi_{\phi}^2}{\eta}
-2\left(\frac m{\eta}+\frac{\cal E}{2}\,\eta\right)\,\Pi_0\,\Pi_{\phi}
-2\left(m+v_0\eta-\frac{\cal E}{2}\eta^2\right)H\\[4mm]\dst\hspace{4cm}
+\left(\frac{m^2}{\eta}+2v_0m+(v_0^2-3m{\cal E})\eta-v_0{\cal E}\eta^2
+\frac{{\cal E}^2}{4}\eta^3\right)\Pi_0^2.\earr\eeq
The relation with the K-S tensor of section 3 is $\ 
{\cal S}_{III}=-\frac 12\,C_{\xi}.$

Having settled the case of an extra holomorphic Killing vector let us now 
consider the case of an extra tri-holomorphic Killing vector.

\section{H-J separability: extra tri-holomorphic Killing}
We write the metric in the form
\beq\label{met2}
g=\frac 1V(dt+G\,dz)^2+V\left(dz^2+\ga_1\,d\xi_1^2+\ga_2\,d\xi_2^2\right),\eeq
where the coordinates $\xi_1$ and $\xi_2$ will be appropriate coordinates in the 
xy-plane which will ensure separability. The two commuting Killing vectors 
$\,\wti{\cal K}=\pt_t\,$ and $\,\wti{\cal L}=\pt_z,$ both tri-holomorphic, are apparent.

The hamiltonian is
\beq\label{ham2}
H=\frac{V^2+G^2}{2V}\,\Pi_0^2-\frac GV\,\Pi_0\,\Pi_z+\frac{\Pi_z^2}{2V}
+\frac 1{2V}\left(\frac{\Pi_1^2}{\ga_1}+\frac{\Pi_2^2}{\ga_2}\right).\eeq
It follows that $\,\Pi_0\,$ and $\,\Pi_z\,$ are conserved.

\subsection{First case}
We use elliptic coordinates $\,\xi_1=\xi\,$ and $\,\xi_2=\eta\,$ in the xy-plane 
defined by
\[x=\frac 1c\,\sqrt{(\xi^2-c^2)(c^2-\eta^2)},\qq y=\frac 1c\,\xi\eta.\]
For convenience, we will define
\[\wh{\xi}=\xi\sqrt{\xi^2-c^2},\qq\wh{\eta}=\eta\sqrt{c^2-\eta^2}.\]
The first case corresponds to
\beq\label{hfc1}
\ga_1=\frac{\xi^2-\eta^2}{\xi^2-c^2},\qq \ga_2=\frac{\xi^2-\eta^2}{c^2-\eta^2},\qq 
 V=v_0+ \frac{a\wh{\xi}+b\wh{\eta}}{\xi^2-\eta^2},\qq
G=\frac{-b\wh{\xi}+a\wh{\eta}}{\xi^2-\eta^2}.\eeq
The separation constants in the H-J equation are
\beq\label{hfc3}\barr{l}
C_{\xi}=(\xi^2-c^2)\Pi_{\xi}^2
+\left(v_0^2\,\xi^2+2v_0a\,\wh{\xi}+(a^2+b^2)(\xi^2-c^2/2)\right)\Pi_0^2
\\[4mm]\hspace{7cm}
+2\,b\,\wh{\xi}\ \Pi_0\,\Pi_z+\xi^2\,\Pi_z^2-2(v_0\,\xi^2+a\,\wh{\xi})\,H,
\earr\eeq
and
\beq\label{hfc4}\barr{l}
C_{\eta}=(c^2-\eta^2)\Pi_{\eta}^2 
+\left(-v_0^2\,\eta^2+2v_0b\,\wh{\eta}+(a^2+b^2)(\eta^2-c^2/2)\right)\Pi_0^2
\\[4mm]\hspace{7cm}
-2\,a\,\wh{\eta}\ \Pi_0\,\Pi_z-\eta^2\,\Pi_z^2+2(v_0\,\eta^2-b\,\wh{\eta})H.
\earr\eeq
The relation with the K-S tensor obtained in section 4 is 
\beq\label{hcf5}
{\cal S}_1=-C_{\xi}+c^2(\Pi_z^2+v_0^2\,\Pi_0^2-2v_0\,H).\eeq

\subsection{Second case}
We use polar coordinates $\,\xi_1=r,\,\xi_2=\phi\,$ in the xy-plane. The second 
case corresponds to
\beq\label{hsc1}
\ga_1=1,\qq \ga_2=r^2,\qq  
V=v_0+m\,\frac{\cos(2\phi)}{r^2},\qq G=-m\,\frac{\sin(2\phi)}{r^2}.\eeq
The separation constants in the H-J equation are
\beq\label{hsc3}\left\{\barr{l}\dst 
C_r=r^2(\Pi_r^2+\Pi_z^2)+\left(v_0^2r^2+\frac{m^2}{r^2}\right)\Pi_0^2-2v_0r^2\,H,
\\[5mm]\dst 
C_{\phi}=\Pi_{\phi}^2+2m\,\sin(2\phi)\,\Pi_0\,\Pi_z+2m\,\cos(2\phi)\left(
v_0\,\Pi_0^2-H\right).\earr\right.\eeq
The relation with the K-S tensor obtained in section 4 is $\ 
{\cal S}_2=C_{\phi}.$

\subsection{Third case}
We use squared parabolic coordinates $\,\xi_1=\xi,\,\xi_2=\eta\,$ in the xy-plane. The 
third case corresponds to
\beq\label{htc1}
\ga_1=\ga_2=\xi^2+\eta^2,\qq 
V=\frac{a\,\xi+b\,\eta}{\xi^2+\eta^2},\qq G=\frac{b\,\xi-a\,\eta}{\xi^2+\eta^2}.
\eeq
The separation constants in the H-J equation are
\beq\label{htc3}\left\{\barr{l}
C_{\xi}=\Pi_{\xi}^2+(\xi\,\Pi_z-b\,\Pi_0)^2
+\frac 12(a^2-b^2)\Pi_0^2-2a\xi\,H,\\[4mm]
C_{\eta}=\Pi_{\eta}^2+(\eta\,\Pi_z+a\eta\,\Pi_0)^2
-\frac 12(a^2-b^2)\Pi_0^2-2b\eta\,H.\earr\right.\eeq
The relation with the K-S tensor obtained in section 4 is $\ 
{\cal S}_3=-\frac 12\,C_{\xi}.$

\subsection{Fourth case}
We use cartesian coordinates $\,\xi_1=x,\,\xi_2=y\,$ in the xy-plane. The 
fourth case corresponds to
\beq\label{hFc1}
\ga_1=\ga_2=1,\qq V=v_0+mx,\qq G=my.\eeq
The separation constants in the H-J equation are 
\beq\label{hFc3}\left\{\barr{l}
C_x=\Pi_x^2+V^2\Pi_0^2-2VH,\\[4mm]
C_y=\Pi_y^2+(\Pi_z-my\,\Pi_0)^2.\earr\right.\eeq
The relation with the K-S tensors obtained in section 4 is merely 
$\ {\cal S}_4^{(1)}=C_y.$

As a conclusion of these last two sections let us observe that the separable 
coordinates, known for the various potentials $V,$ lift up, without any modification,  
to separable coordinates for the four dimensional system. Let us turn now to the 
Killing-Yano tensors.

\section{Killing-Yano tensors}
An antisymmetric tensor $Y_{\mu\nu}$ is a Killing-Yano (K-Y) tensor iff
\beq\label{ky1}
\nabla_{(\mu}\,Y_{\nu)\rho}=0.\eeq
A complex structure is therefore a K-Y tensor.

The usefulness of such a concept is related to the fact that the symmetrized 
tensor product of two K-Y tensors does give a K-S tensor, as can be checked by an 
easy computation. Clearly the triplet of complex structures shared by the multi-centre 
metrics is not very useful since it gives only trivial K-S tensors so we need extra 
K-Y tensors. It is the aim of this section to give new examples of these extra K-Y 
tensors which will give some explicit K-S tensors which do not satisfy assumption (A 3).

We have been able to obtain K-Y tensors for 
\brm
\item The special case of the second dipolar breaking, corresponding to 
$\,V=v_0+{\cal E}\,z.$
\item The fourth case with an extra tri-holomorphic Killing vector, with potential 
$\,V=v_0+m\,x.$
\erm
Let us consider successively these two cases.

\subsection{Special second dipolar breaking}
For $m=0$ the metric simplifies to
\beq\label{metsp1}
g=\frac 1{4V}\left(2dt-{\cal E}y\,dx+{\cal E}x\,dy\right)^2+V(dx^2+dy^2+dz^2),
\qq V=v_0+{\cal E}z.\eeq
We have four Killing vectors 
\beq\label{msp2}
\pt_t,\quad x\,\pt_y-y\,\pt_x,\quad \pt_x+\frac{{\cal E}y}{2}\,\pt_t,
\quad \pt_y-\frac{{\cal E}x}{2}\,\pt_t,\eeq
and the induced conserved quantities have simple Poisson brackets: $\,\Pi_0\,$ is 
central and for the remaining ones
\beq\label{msp3}
\{J_z,p_x \}=p_y ,\qq 
\{J_z,p_y\}=-p_x\,\qq 
\{p_x,p_y\}={\cal E}\,\Pi_0.
\eeq
with
\[J_z=x\,\Pi_y-y\,\Pi_x,\qq p_x=\Pi_x+\frac{{\cal E}y}{2}\,\Pi_0,
\qq p_y=\Pi_y-\frac{ {\cal E}x}{2}\,\Pi_0.\]

Using the canonical vierbein one gets for the K-Y two-form 
\beq\label{metsp4}
Y=-{\cal E}^2\,E_0\wedge(x\,E_1+y\,E_2)+{\cal E}^2(x\,E_2\wedge E_3+y\,E_3\wedge E_1)
+2{\cal E}V\,E_1\wedge E_2.\eeq
\ From it and the complex structures we can construct four K-S tensors
\[Y^2,\qq \qq S_i=Y\,\Om^{(-)}_i+\Om^{(-)}_i\,Y,\qq i=1,2,3\]
We will quote the corresponding conserved quantities instead of the K-S tensors, for 
the ease of comparison with our earlier results:
\beq\label{metsp5}\barr{l}\dst 
\frac{Y^2}{{\cal E}^2}\ \rightarrow\ -4V(\Pi_x^2+\Pi_y^2)+{\cal E}^2(x^2+y^2)V\,\Pi_0^2
+4{\cal E}\,\Pi_{\phi}(x\,\Pi_x+y\,\Pi_y)-2{\cal E}^2(x^2+y^2)H,\\[4mm] 
S_1\ \rightarrow\ 4{\cal E}V\,\Pi_0\,p_y
-4{\cal E}\,\Pi_{\phi}\,p_x+4{\cal E}^2x\,H,\\[4mm] 
S_2\ \rightarrow\  -4{\cal E}V\,\Pi_0\,p_x
-4{\cal E}\,\Pi_{\phi}\,p_y+4{\cal E}^2y\,H,\\[4mm]
S_3\ \rightarrow\ 4{\cal E}(p_x^2+p_y^2).\earr\eeq
Let us observe that $\,S_3\,$ is {\em reducible} and that $\,S_1\,$ and $\,S_2\,$ 
do not satisfy (A 3), so we are left with $Y^2.$ Some algebra shows how it 
is related to the conserved quantity obtained in section 3:
\beq\label{metsp6}
{\cal S}_{III}(m=0)=-\frac{Y^2}{4{\cal E}^3}-
\frac{v_0}{4{\cal E}^2}\,S_3-v_0\,\Pi_0\,J_z,\eeq
so that, up to reducible terms, the two conserved quantities are one and the same. 
This case is quite similar to the Kerr metric (albeit much simpler) for which the 
Carter K-S tensor is in fact the square of some K-Y tensor.

\subsection{The fourth case}
Using the canonical vierbein one gets for the K-Y two-form 
\beq\label{yfc1}
Y=-my\,\Om^{(-)}_2-mz\,\Om^{(-)}_3+2V\,E_2\wedge E_3.\eeq
Defining $\,p_z=\Pi_z-G\,\Pi_0,$ we can write the induced conserved quantities:
\beq\label{yfc2}\barr{l}
\barr{l}\dst 
\frac{Y^2}{4}\ \rightarrow\ -V\,\Pi_y^2-V\,p_z^2
+my\,\Pi_x\,\Pi_y-myV\,\Pi_0\,p_z\\[2mm]\hspace{4cm}
+mz\,\Pi_x\,p_z+mzV\,\Pi_0\,\Pi_y-\frac{m^2}{2}(y^2+z^2)H,\earr\\[7mm]
S_1\ \rightarrow\ \Pi_y^2+p_z^2,\qq p_z=\Pi_z-G\,\Pi_0,\\[4mm]\dst 
\frac{S_2}{4}\ \rightarrow\ -\Pi_x\,\Pi_y+V\,\Pi_0\,p_z+my\,H,\\[4mm]\dst 
\frac{S_3}{4}\ \rightarrow\ -\Pi_x\,p_z-V\,\Pi_0\,\Pi_y+mz\,H.\earr\eeq
We see that $\,S_1\,$ and $\,S_2\,$ were alredy obtained in section 4. The other two 
are missing since they don't satisfy our assumption (A 3). Notice 
also that the conserved quantity $\,{\cal S}_4^{(2)}\,$ cannot be obtained in that way.

So this example is of some interest since it shows that there do exist K-S tensors 
which do not satisfy the assumption (A 3). However, since the corresponding 
conserved quantities do not commute with $\,\Pi_z,$ they are of no use to prove 
integrability.

\section{Conclusion}
We have settled the problem of finding all the multi-centre metrics which do 
exhibit some extra conserved quantity, having the structure (\ref{ks4}),  
under the assumptions (A 1) to (A 3). Since it is 
induced by a KS tensor, this conserved quantity is quadratic with respect to 
the momenta, and preserved by the geodesic flow. As we have observed, the existence 
of this extra conserved quantity is essential to obtain integrability. 

However one should keep present that our analysis does not cover all the 
{\em integrable} multi-centre metrics, since integrability could emerge from 
the existence of more complicated conserved quantities. In fact the concept 
of Killing-St\" ackel tensor can be generalized to symmetric $\,(n,0)\,$ tensors with 
$\,n\geq 3$ such that 
\[\nabla_{(\la}S_{\mu_1\cdots\mu_n)}=0.\]
It follows that the geodesic flow preserves the quantity
\[S_{\mu_1\cdots\mu_n}\,\dx^{\mu_1}\cdots\dx^{\mu_n}.\]
The corresponding invariants will be cubic, quartic, etc... with respect to 
the momenta. Little is known about the existence of such 
conserved quantities, which could produce possibly new integrable 
multi-centre metrics. 

Let us conclude by putting some emphasis on the purely local nature of our 
analysis: it makes no difference between complete and non-complete metrics. For 
instance in section 4 we have seen that the most general two-centre metric 
is integrable, however it is complete only for real $\,m_1=m_2,$ i. e. for 
the double Taub-NUT metric.



\end{document}